\documentclass[5p]{elsarticle}

\usepackage{graphicx}

\usepackage{amssymb}

\journal{Nuclear Instruments and Methods A}

\begin{document}

\begin{frontmatter}



\title{Silicon photomultiplier timing performance study}


\author[A,B]{Ahmed Gamal\corref{C}}
\author[A]{B\"uhler Paul} 
\author[A]{Marton Johann}
\author[A]{Suzuki Ken}
\address[A]{Stefan Meyer Institute for Subatomic Physics of the Austrian Academy
of Sciences, Vienna, Austria}
\address[B]{Al-Azhar University, Faculty of Science, Physics Department, Cairo,
Egypt}
\cortext[C]{Corresponding author: gamal.ahmed@assoc.oeaw.ac.at}

\begin{abstract}

Many characteristics of Silicon Photomultipliers can be tuned with temperature
and operation voltage. We present preliminary results of a study of the effect
of these two operation parameters on the time resolution of large active area
Multi-Pixel Photon Counters (MPPCs) manufactured by Hamamatsu. Measurements at
$-10$~$^{\circ}$C, $0$~$^{\circ}$C, and $10$~$^{\circ}$C at different bias voltages
were performed. The time resolution is best at low temperature and high
over-voltage. Most significant improvements can be achieved in cases with low
number of fired pixels ($<$10 pixels). Between the worst and best case among the
considered conditions a factor of 5 improvement was observed. In cases with
large number of fired pixels ($>$40 pixels) the effect of temperature and
operation voltage becomes smaller. The timing performance still improves with
decreasing temperature ($\approx$ factor of 2) but it hardly depends on the
operation voltage. The study shows, that especially in applications where only
few photons are available for detection a careful optimization of temperature
and operation voltage are advisable to obtain optimum timing results with the
MPPC.

\end{abstract}


\end{frontmatter}


\section{Introduction}

SiPMs have numerous advantages as compared to other photo detectors. The low
operation voltage, high gain, fast timing, compactness, and moreover the
insensitivity to magnetic fields make them excellent candidates for different
applications \cite{buzhan03a,suzuki09a,ahmed10a}. In many applications in
experimental physics time information is gained from light detecting devices. In
subatomic physics e.g. time-of-flight (TOF) measurements, using scintillation
counters in combination with light sensitive devices is a widely used technique
to determine the velocity of a particle and in combination with other detectors
to perform particle identification (PID). The performance of such detector
systems crucially depends on the time resolution of the photo sensor used. In
PID a better time resolution of the photo sensor allows better particle
separation. The operation characteristics of SiPM depends mainly on two
parameters - the operation voltage and temperature. In view of possible usage of
SiPMs in timing applications we investigate its timing performance as function
of these two parameters. 

For the current study we focus on the recent large active area SiPMs - the
Multi-Pixel Photon Counters (MPPCs) from Hamamatsu. These MPPC photo sensors are
sensitive in the blue-light range ($\approx 400$~nm) which matches well with the
light emitted by scintillators and Cherenkov radiators. The devices of the
series S10362-33-100C \cite{hamamatsu} which we are investigating, have a
sensitive area of 3$\times$3 mm$^2$. They consist of 900 APDs of 100$\times$100
$\mu m^2$ with a total fill factor of about 78.5\%. Their nominal gain is
$\approx 2.4\times 10^6$ and they have a terminal capacitance of 320~pF. The
photon detection efficiency in the blue light range depends on temperature and
operation voltage and is typically $\ge 50$\% . According to Hamamatsu the time
resolution at $25$~$^{\circ}$C, nominal operation voltage, and single photon level
is $\sigma \approx 230$~ps. Measurements of dark current, dark count rate and
cross-talk of this device have been presented elsewhere
\cite{renker06a,ahmed09a}.

\section{Measurement setup}

\begin{figure}[hbt] 
\centering 
\includegraphics[width=0.5\textwidth,keepaspectratio]{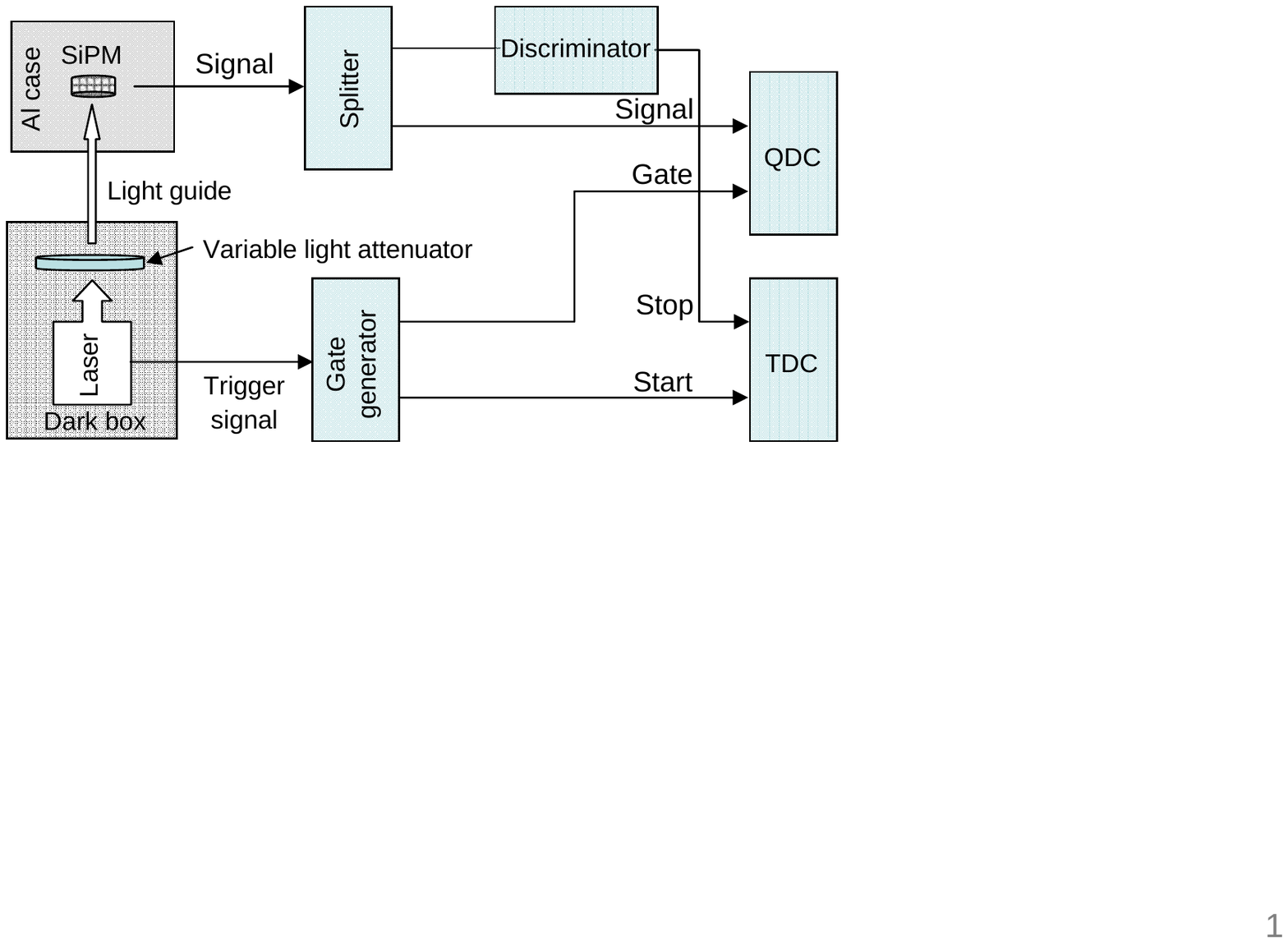}
\caption{Sketch of the set-up for the SiPM timing study. The MPPC, which is
placed in a light tight box is stimulated with short light pulses from a
picosecond laser. The data acquisition electronics is setup to measure charge
and time distributions simultaneously (see text for details).}
\label{fig01}
\end{figure}

The setup used to study the timing performance of the MPPC device is sketched in
figure \ref{fig01}. 

To illuminate the SiPM with blue laser light, we use a picosecond blue laser
system from Advanced Laser Diode Systems (PIL063SM) \cite{alds} equipped with a
408 nm head with a pulse width of $<45$ ps. The light intensity on the SiPM is
controlled by the tunable laser power and an attenuation filter. The repetition
rate can be regulated from single shot to 1~MHz. The attenuated light was
delivered to the SiPM by an optical fiber of 1~mm diameter.

The SiPM housing is mounted in a light and vacuum tight aluminum box. The SiPMs
are thermally coupled to water-cooled peltier elements, which allows to regulate
the temperature of the photo sensor from room temperature down to approximately
$-20$~$^{\circ}$C with $\pm 0.1$~$^{\circ}$C of accuracy. In order to avoid
condensation, the aluminum box is evacuated to a pressure of $\approx
10^{-3}$~mbar. The SiPM signal is amplified with a fast preamplifier (AMP\_0611)
from Photonique SA \cite{photonique}. In order to minimize the electronic noise
pick-up the SiPMs are attached directly to the preamplifier board.

The readout electronics is set up to measure charge distributions and time
distributions, simultaneously. The signal output from the SiPM is split into two
lines. One is connected to a charge-to-digital converter (QDC, LeCroy ADC-2249W,
$0.25$~pC/ch) for the charge measurements. The other line is fed into a leading
edge discriminator and the output of the discriminator is fed into a
time-to-digital converter (TDC, Phillips TDC-7186, $25$~ps/ch) for time
measurements (stop signal). The data acquisition is triggered by the trigger-out
signal of the laser. It is used as start for the TDC and also to generate a
sampling gate for the QDC.

The QDC and TDC signals were recorded by a personal computer via Wiener
CAMAC-CC32 PCI bus interface and stored for offline analysis.

\section{Measurements and results}

With this series of measurements we aim at investigating the temperature and
bias voltage dependence of the timing performance of the MPPC. Therefore we
determined the time resolution at three different temperatures ($10$~$^{\circ}$C,
$0$~$^{\circ}$C, $-10$~$^{\circ}$C) and several bias voltages as function of the
number of fired pixels. The measurement were performed at totally 13 pairs of
temperature/bias voltage values.

The input light intensity to the MPPC was adjusted at the beginning of the
measurement series and was kept constant for all measurements. The pulse
frequency was set to $1$~kHz.

Instead of presenting the data as function of bias voltage we use the
over-voltage as parameter. The over-voltage is defined as the difference between
the applied bias voltage and the breakdown voltage, which is a function of
temperature. The breakdown voltage was determined as described in
\cite{vinke09a}. For the specific example of measured MPPC the breakdown voltage
is $68.58\pm 0.05$~V at $10$~$^{\circ}$C, $67.92\pm 0.05$~V at $0$~$^{\circ}$C,
$67.45\pm 0.05$~V at $-10$~$^{\circ}$C.

\begin{figure}[hbt] 
\centering 
\includegraphics[height=0.5\textwidth,keepaspectratio,angle=90]{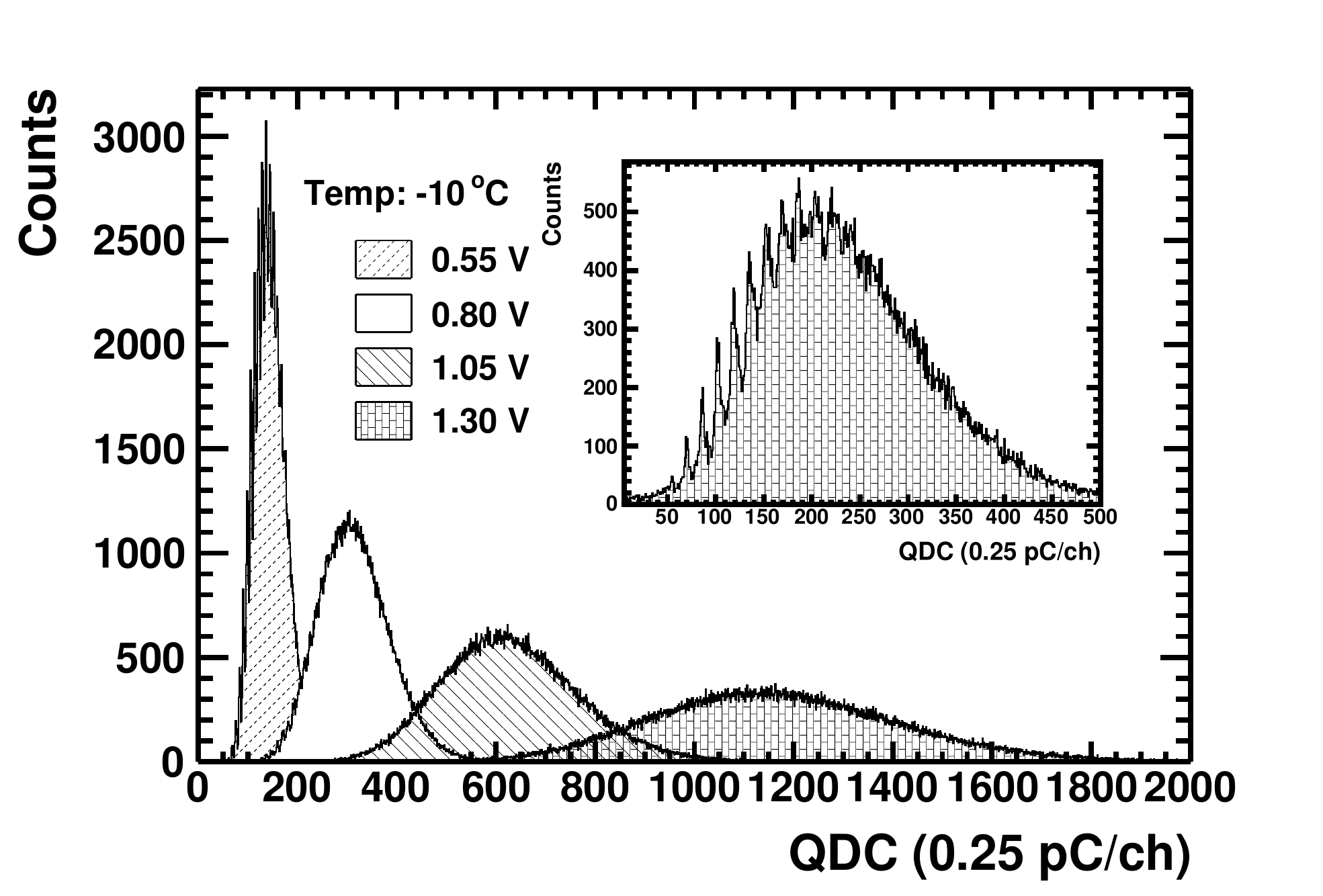}
\caption{Charge distributions of the MPPC S10362-33-100C at $-10$~$^{\circ}$C and
different over-voltages. The inset shows the distribution acquired at an
over-voltage of $1.3$~V but at lower light intensity than the distribution
below. In this case single photon peaks are resolved which can be explored to
determine the gain of the device at the given operation condition.}
\label{fig02}
\end{figure}

For each pair of temperature and bias voltage the measurement procedure was as
follows.

\begin{itemize}
\item Determine gain and conversion factor between QDC value and number of fired
pixels.

Both quantities can be determined from the analysis of separated peaks
in the charge distribution corresponding to different values of fired pixels. An
example of such a distribution is displayed in the inset of figure \ref{fig02}.
QDC value and number of fired pixels (NFP) are related by

\begin{equation}
NFP = \frac{QDC-QDC_1}{fac} + 1
\label{equ01}
\end{equation}

where $QDC_j$ is the position of the peak corresponding to $j$ fired pixel and
$fac=\frac{QDC_m-QDC_n}{m-n}$. The gain G is given by

\begin{equation}
G = \frac{fac\cdot 0.25\cdot 10^{-12}}{e\cdot G_{pa}}
\label{equ02}
\end{equation}

where $e$ the charge of an electron is $1.602\cdot 10^{-19}$~C and the
gain of the preamplifier $G_{pa}=4.7$.

In most cases to be able to see separate single photon peaks in the charge
distribution the light intensity needed to be attenuated. For the timing
measurement, however, the light intensity was reset to the nominal value. E.g.
the most right charge distribution in figure \ref{fig02} and the distribution in
the inset of figure \ref{fig02} have been taken at the same temperature and
over-voltage but at different light intensities. Whereas the first, taken at
nominal light intensity, is used for timing measurements the later is used to
determine the gain at the given operation conditions.

\item Select in QDC spectrum narrow ranges corresponding to a given number of
fired pixels.

The charge distributions measured at nominal light intensity are broad as shown
in figure \ref{fig02}. In order to determine the time resolution as function of
NFP we select narrow ranges of the charge distributions which can be associated
with a given number of fired pixels through relation (\ref{equ01}).

With this procedure the time resolution dependence on the NFP is obtained by
selecting successive narrow portions (corresponding to 1 or 2 fired pixels in
width) of a broad distribution of NFP and not by tuning the input light
intensity. The range of NFP which can be grasped with this method is limited and
different for each SiPM operation condition. This is because the charge
distributions only cover a portion of the full QDC range, which depends on the
SiPM operation condition (see figure \ref{fig02}). In order to gain access to a
broader range of NFP, the light intensity needed to be changed. This we
explicitly excluded for this study not to bias the results by a possible
influence on the incident light intensity.

\item Apply slew correction and fit the resulting TDC distributions with
a Gaussian function.

Since we are using a leading edge discriminator in the timing channel a slew
correction is applied to correct for the signal height dependence of the stop
signal. The slew correction parameters are determined with all data points. The
corrected TDC distribution of each NFP-bin is fitted with a Gaussian function.
The time resolution is defined to be the sigma of the fitted width. Since the
NFP-bins are narrow the slew correction has only a small effect on the time
resolution of an individual bin.

\end{itemize}

Changing and stabilizing the temperature of the system takes much longer than
changing the bias voltage. Therefore we first performed the measurements at
fixed temperature and different bias voltages and then stepped in temperature.

\begin{figure}[hbt] 
\centering 
\includegraphics[width=0.5\textwidth,keepaspectratio]{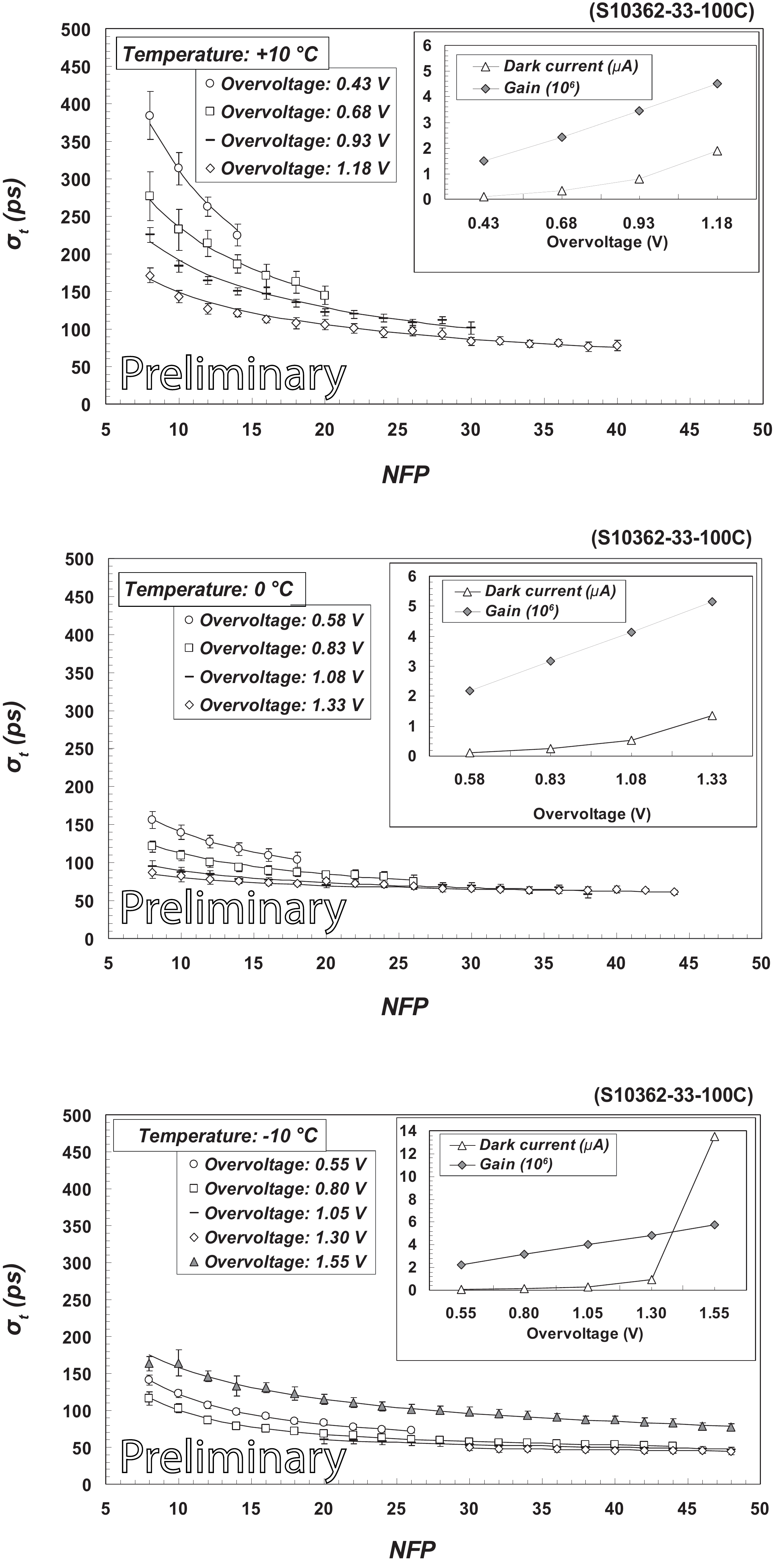}

\caption{Time resolution as function of NFP of the MPPC S10362-33-100C at
$10$~$^{\circ}$C (upper panel), $0$~$^{\circ}$C (middle panel), and $-10$~$^{\circ}$C
(lower panel) and different over-voltages. The insets show gain and dark current
as function of over-voltage. Time resolution is a strong function of temperature
and over-voltage. Especially at low NFP a proper tuning of the operation
conditions is required to obtain optimal timimg performance.}
\label{fig04}
\end{figure}

The results of these measurements are summarized in figure \ref{fig04}. The
panels show results for $10$~$^{\circ}$C (upper panel), $0$~$^{\circ}$C (middle
panel), and $-10$~$^{\circ}$C (lower panel). The time resolution is plotted as NFP
for different values of over-voltage. In the insets the gain and dark current
are displayed as function of over-voltage.

The contribution from the data acquisition electronics to the time resolution
was measured by feeding a constant-amplitude test-pulse into the system. The
contribution of the electronics measured in this way (excluding contributions
from the preamplifier) is $\approx$ 20 ps.

\section{Discussion and summary}

The measurements of the time resolution $\sigma_t$ of the MPPC at different
temperatures and over-voltages presented in the previous section reveal three
general trends - 1. the time resolution improves with increasing NFP, 2. within
the covered temperature range of $-10$~$^{\circ}$C and $10$~$^{\circ}$C the time
resolution improves with decreasing temperature, and 3. the time resolution
improves with increasing over-voltage (except for extreme cases which are
discussed below).

In order to guide the eye the $\sigma_t(NFP)$ curves at given temperature and
over-voltage are fitted with the function $\sigma_t(NFP) = A + B/\sqrt{NFP}$.
The results are shown as bold lines in figure \ref{fig04}. At first view the
function seems to fit well the data. However, closer inspection reveals
deviations of the data points at low NFP from this relation. This needs further
investigation.


Temperature and over-voltage dependence of $\sigma_t$ are most prominent at
small NFP. Whereas at $10$ fired pixels the range of $\sigma_t$ values measured
at the selected conditions is from $\approx 80$ to $400$~ps (factor~5) the range
at $40$ fired pixels is much smaller (only from $\approx 40$ to $80$~ps,
factor~2). The temperature dependence is not linear. The improvement of
$\sigma_t$ which is achieved by decreasing the temperature from $10$~$^{\circ}$C to
$0$~$^{\circ}$C is considerably larger than the improvement achieved by lowering
the temperature by additional $10$~$^{\circ}$C to $-10$~$^{\circ}$C. How much the time
resolution can be improved by choosing the appropriate temperature also depends
on the over-voltage applied to the MPPC. The influence of the temperature is
large when the MPPC is operated with a low over-voltage. With increasing
over-voltage the importance of the temperature decreases.

Although operating the MPPC at large over-voltage has a positive effect on the
timing performance this general rule breaks down when the bias voltage is
substantially riced above the recommanded operation range. Then the dark current
starts to steeply increase and a deterioration of the time resolution is
observed. An example of this behavior can be observed in the $-10$~$^{\circ}$C-data
at an over-voltage of $1.55$~V (lower panel of figure \ref{fig04}). A similar
behaviour is also observed in other types of MPPCs (see e.g. figure 4 in
\cite{ronzhin10a}). Operating the MPPC at large over-voltage is accompanied by
an increase of the dark current and dark counts (see insets in figure
\ref{fig04}). This can be a severe drawback in applications where only few
photons are available for detection.

To obtain best performance of the MPPC for a given application a careful
optimization of the operation conditions is obviously required. To optimize
timing performance and also to achieve low dark count rates, control of the
temperature is most important. Especially in applications with low light
intensities (e.g. Cerenkov counters) the timing performance can be significantly
improved by operating the device at low temperatures. Our current measurements
reach down to $-10$~$^{\circ}$C. A further improvement of the timing performance
can be expected at even lower temperatures \cite{collazuol10a}. Using high
over-voltage can also improve the time resolution. This is however accompanied
with an increase of the dark counts. Attention must be paid not to operate the
MPPC at an over-voltage above the value at which the dark current starts to
rapidly increase and the timing performance starts to degrade. An optimum
operation voltage seems to be slightly below this reversal point. However, since
the over-voltage corresponding to the reversal point strongly depends on the
temperature (see e.g. figure 1 in \cite{ahmed10a}) operation at this point
requires stable temperature control.


\section{Acknowledgments}

This work is partly supported by INTAS (project 05-1000008-8114) and
Hadronphysics2 (project 227431). One of us (G.A.) acknowledges the support by
the Egyptian Ministry of higher education.




\end{document}